# Switchable and unswitchable bulk photovoltaic effect in two-dimensional interlayer-sliding ferroelectrics

Rui-Chun Xiao[1], Yang Gao[2,*], Hua Jiang[3], Wei Gan[1], Changjin Zhang[1,4], Hui Li[1,*]

[1]*Institute of Physical Science and Information Technology and Information Materials and Intelligent Sensing Laboratory of Anhui Province, Anhui University, Hefei 230601, China*

[2]*Department of Physics, University of Science and Technology of China, Hefei 230026, China*

[3]*School of Physical Science and Technology, Soochow University, Suzhou 215006, China*

[4]*High Magnetic Field Laboratory, Chinese Academy of Sciences, Hefei 230031, China*

Email: ygao87@ustc.edu.cn (Y. G.), huili@ahu.edu.cn (H. L.)

## Abstract

Spontaneous polarization and bulk photovoltaic effect (BPVE) are two concomitant physical properties in ferroelectric materials. The flipping of ferroelectric order usually accompanies with the switching of BPVE as both of them are reversed under the inversion symmetry. In this study, we report the distinctive BPVE characters in two-dimensional (2D) interlayer sliding ferroelectric materials featuring unswitchable in-plane BPVE (light-induced photocurrent in the *xy* plane) and switchable out-of-plane BPVE (light-induced polarization along the *z*-direction). Symmetry analysis within abstract bilayer crystal model and first-principles calculations validate these BPVE properties. It is because the positive and negative ferroelectric states caused by interlayer sliding are related by mirror symmetry which cannot flip all the BPVE tensor elements. This finding extends the understanding of the relationship between ferroelectricity and BPVE. On one hand, the switchable out-of-plane BPVE can be used to design switchable photoelectric devices. On the other hand, the in-plane BPVE is robust against the ferroelectric flipping, and the unswitchable character is beneficial to construct larger-scale photoelectric devices.

## Introduction

Ferroelectricity and bulk photovoltaic effect (BPVE)[1,2] are two basic physical phenomena that emerge in condensed matter materials due to symmetry breaking. Generally speaking, BPVE, an intrinsic optical rectification phenomenon, appears in materials without inversion symmetry which can convert light to electricity as solar cells, hence is naturally fulfilled the symmetry requirement in ferroelectric materials. As a result, the ferroelectric materials with switchable electronic polarization are one of the most studied BPVE materials[1-3], and the corresponding energy conversion efficiency can exceed the Shockley-Queisser limit[3-5].

Under the inversion operation, both ferroelectric order and BPVE switch sign. Therefore, the ferroelectric polarization is often taken as a handle to manipulate the direction of BPVE photocurrent in, e.g., perovskite oxides $BiFeO_3$[6-8], $Pt/BiFeO_3/SrRuO_3$[9], 2D (two-dimensional) $CuInP_2Se_6$[10], charge-transfer complex[11], SnTe monolayer[12], *etc.*. However, from the aspect of symmetry, the ferroelectric polarization is a polar vector, while the BPVE coefficients form a rank-three tensor. They are hence subject to distinct symmetry transformation rules, and should not always synchronize. For example, a mirror reflection perpendicular to the polarization will reverse the polarization just as the inversion operation does, while it cannot flip all the components of the BPVE. Thus, precise understanding of the relationship between the ferroelectric order and the BPVE is of tremendous significance in both fundamental research and the design of photoelectric devices.

In this work, we think the exceptional case that BPVE and ferroelectric order are in-step can be found in 2D interlayer-sliding ferroelectric materials[13]. As demonstrated in recent experiments, the out-of-plane ferroelectricity can be achieved in bilayer $WTe_2$[14-17], BN[18,19], InSe[20,21], $MoS_2$[22,23] *etc.* by interlayer sliding, which effectively extends the family of 2D ferroelectric materials[24-27]. This unique ferroelectric flip mechanism may lead to the BPVE property beyond the switchable scenario.

Focusing on four most common stacking cases of 2D bilayers with interlayer sliding, we use an *abstract bilayer crystal model* to locate the available configurations and symmetries with nontrivial ferroelectric order. We find two cases (Case 1b and Case 2a) can have the possibility to achieve the interlayer-sliding ferroelectricity, moreover the opposite ferroelectric states in the two cases are all related by the mirror operator. Coincidentally, the interlayer-sliding ferroelectric materials found in the experiments are attributed to these two cases. Using bilayer $MoS_2$ and $WTe_2$ as representative examples, we performed first-principles calculations, and find that the in-plane BPVE does not change the sign with ferroelectric order, which is different from conventional ferroelectrics. In contrast to in-plane BPVE, out-of-plane BPVE switches with the reverse of the ferroelectric order. These characters will lead to distinct photovoltaic phenomena in experiments.

# Polarization flipping via interlayer sliding

Now we introduce the mechanism of the polarization flipping via interlayer sliding. **Figure 1**(a) shows the out-of-plane polarization $P_z$ in bilayer vdW (van der Waals) material without inversion nor horizontal mirror symmetry. Because the in-plane polarization is usually annihilated by in-plane rotational symmetry in real materials, the $+P_z$ and $-P_z$ polarization are denoted as $+P$ and $-P$ for short. The out-of-plane polarization can be flipped in two different ways: (i) intralayer movement, *i.e.*, atomic relative displacements in each monolayer [**Fig. 1**(b)], and (ii) interlayer sliding[13] where the crystal structure of each monolayer is invariant [**Fig. 1**(c)]. Although polarization flipping can be achieved in both situations, the underlying symmetries of the switch of ferroelectric states are different. For situation (i), the $+P$ and $-P$ ferroelectric states are correlated by the inversion operator, as indicated in unit cells in **Fig. 1**(a) and **Fig. 1**(b). While for situation (ii), the $+P$ and $-P$ ferroelectric states are correlated by the mirror symmetry $\widehat{M}_{xy}$, as seen in unit cells in **Fig. 1**(a) and **Fig. 1**(c) (detailed analyses are presented in the Symmetry analysis section for real materials). The interlayer sliding mechanism is feasible experimentally, and has been observed in bilayer vdW materials[14-16,18-23] as stated above, which further extends the family of 2D ferroelectric materials[24-27] effectively. **Figure 1**(d)-(f) shows the basis vectors transformation under the inversion symmetry and mirror symmetry. The BPVE in these interlayer-sliding ferroelectric materials should not always change with the ferroelectric order.

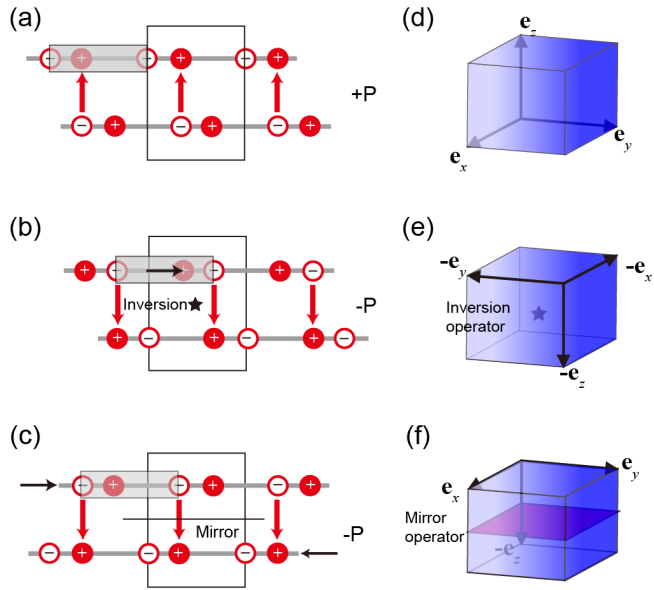

**Fig. 1 Polarization flipping in a 2D vdW bilayer material via intralayer movement or interlayer sliding, and related symmetries.** (a) A 2D system with positive out-of-plane

polarization +P. (b) The polarization is reversed to –P by atom displacements in each monolayer, and the black arrow means the movement of the cations relative to the anions. (c) The polarization is reversed by interlayer sliding, and the black arrows mean the interlayer sliding direction. (d) Three basis vectors and their transformations under (e) inversion symmetry (f) horizontal mirror symmetry. The black rectangles denote the bilayer unit cells in (a)-(c), and the gray rectangles denote the monolayer unit cells.

## Symmetry analysis

Next, we analyze how to design the bilayer vdW materials with the out-of-plane polarization via interlayer sliding in **Fig. 1**(a), and analyze how positive and negative polarizations are related by symmetry operator. Most of the discovered monolayer materials are non-polar and non-ferroelectric (such as BN, transition-metal dichalcogenides, metal trihalide, post-transition metal chalcogenides), and their monolayer crystal structures belong to the following two different cases, as shown in Fig. S1 and Fig. S2 in the Supplementary Information[28,29]:

1. Monolayer has the inversion symmetry $\hat{I}$, but has no horizontal mirror symmetry $\widehat{M}_{xy}$, such as monolayer $1T$-$PtS_2$, $1T'$-$WTe_2$, $BiI_3$, *etc.*;

2. Monolayer has no inversion symmetry $\hat{I}$ but has horizontal mirror symmetry $\widehat{M}_{xy}$, such as monolayer BN, $1H$-$MoS_2$, $1H$-$WSe_2$, GaSe, InSe, *etc.*

Besides, the bilayer can be further divided into two subcases according to the stacking manner of two layers:

(a) A/A stacking, *i.e.* two adjacent layers are identical where the A represents the crystal structure of the monolayer.

(b) A/B stacking, *i.e.* two adjacent layers are relatively rotated by 180° with $B = \hat{C}_{2z}A$ (A and B mean the crystal types).

The above considerations can form four cases for 2D bilayer materials with interlayer sliding, and we can analyze their symmetries with an *abstract bilayer crystal model* which only considers the above crystal symmetries and ignores the specific crystal structures. The complete symmetry analysis for each case is summarized in Supplementary Note 1, and the corresponding results are shown in **Table I**. As can be seen, Case 1b and Case 2a host the out-of-plane polarization, and the two opposite interlayer-sliding states are correlated by $\widehat{M}_{xy}$ operator, instead of inversion $\hat{I}$ operator. In contrast, the crystal structures in Case 1a and Case 2b possess the inversion symmetry whatever the interlayer-sliding vector, which forbids the occurrence of ferroelectricity and BPVE.

**Table I Symmetries of interlayer-sliding bilayer materials.**

| Stacking way / Monolayer symmetry | a. A/A | b. A/B |
|---|---|---|
| 1. $\begin{cases} \hat{I}: & \checkmark \\ \widehat{M}_{xy} & \times \end{cases}$ | 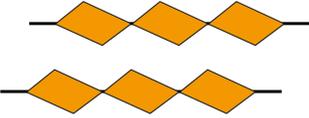 Case 1a: $\begin{cases} \hat{I}: & \checkmark \\ \widehat{M}_{xy} & \times \end{cases}$ | 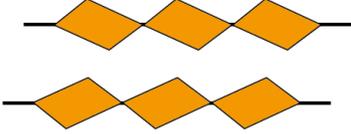 Case 1b: $\begin{cases} \hat{I}: & \times \\ \widehat{M}_{xy}: & \times \\ \widehat{M}_{xy}S_+ = S_- \end{cases}$ |
| 2. $\begin{cases} \hat{I}: & \times \\ \widehat{M}_{xy} & \checkmark \end{cases}$ | 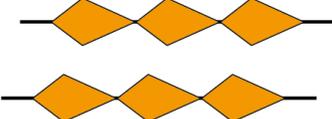 Case 2a: $\begin{cases} \hat{I}: & \times \\ \widehat{M}_{xy}: & \times \\ \widehat{M}_{xy}S_+ = S_- \end{cases}$ | 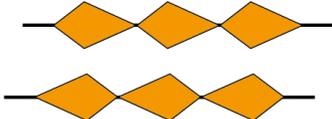 Case 2b: $\begin{cases} \hat{I}: & \checkmark \\ \widehat{M}_{xy}: & \times \end{cases}$ |

Note: (1) The crystal structures here are abstract bilayers, and the orange quadrilateral indicates abstract unit cell of monolayer; (2) $S_+$ and $S_-$ represent the bilayer with opposite sliding vectors.

Table II summarizes several bilayer candidates for the above four cases that can be experimentally realized by either simply mechanical exfoliation or "tear and stack" methods. In particular, the out-of-plane ferroelectricity of the bilayer WTe$_2$[14-17] with Case 1b and BN[18,19], InSe[20,21], bilayer-TMD[22,23] with Case 2a have been discovered in experiments recently.

Table II Candidate bilayer materials and corresponding symmetries of bulk parents

| | Candidate materials | Symmetry of bulk parents |
|---|---|---|
| Case 1a | 1T-MX$_2$ (PtSe$_2$, PtS$_2$, TiS$_2$, SnS2) | $P\bar{3}m1$ (164) |
| | TX$_3$ (BiI$_3$) | $R\bar{3}$ (148) |
| Case 1b | T$_d$/1T'-WTe$_2$, MoTe$_2$, ZrI$_2$ | $Pmn2_1$ (31) or $P2_1m$ (11) |
| Case 2a | GaSe, InSe, InS | $P\bar{6}m2$ (187) |
| | 3R-MX$_2$ (MoS$_2$, MoSe$_2$) | R3m (160) |
| | BN | R3m (160) |
| Case 2b | h-BN | P6$_3$/mmc (194) |
| | 2H-MX$_2$ (MoS$_2$) GaS | P6$_3$/mmc (194) |
| | GeSe, SnSe | Pnma (62) |

Note: (1) MX$_2$ means the transition metal dichalcogenides, and MX$_3$ means the transition metal trihalides, (2) bulk parents mean the bulk materials which staking with the same manners, (3) the structure data are from Materials Project[30].

Now we analyze the characters of in-plane and out-of-plane BPVE of the interlayer-sliding ferroelectric materials in Case 1b and Case 2a under the normal incidence of light. The in-plane BPVE current density is

$$J_a = \sum_{bc} \sigma_{bc}^a E_b(\omega) E_c(-\omega), \tag{1}$$

where $a,b,c \in \{x,y\}$, $E_b$ and $E_c$ are electric fields of light along $b$ and $c$ direction, $\omega$ is the frequency of light, and $\sigma_{bc}^a$ is the BPVE coefficient. In contrast to $x$ and $y$ directions, a static electric polarization instead of the current is formed along the $z$-direction because the $z$-direction of 2D materials is discontinuous. The out-of-plane BPVE polarization is[31]

$$p_z = \sum_{bc} \sigma_{bc}'^z E_b(\omega) E_c(-\omega), \tag{2}$$

where $p_z$ is $z$-component polarizability.

Similar to other nonlinear optics tensors, the BPVE tensor $\left[\sigma_{ab}^c\right]_{3\times3\times3}$ in Eq. (1) [or Eq. (2)] obeys

$$\sigma_{jk}^i = \sum_{abc} R_{ia} R_{jb} R_{kc} \sigma_{ab}^c, \tag{3}$$

where $[R_{ia}]$ is the 3×3 matrix of the symmetry operator $\hat{R}$. As shown in **Fig. 1**(f), $\widehat{M}_{xy}$ will result in the reverse of the out-of-plane vector with invariant in-plane vectors, and the corresponding operator matrix is

$$[M_{xy}] = \begin{bmatrix} 1 & 0 & 0 \\ 0 & 1 & 0 \\ 0 & 0 & -1 \end{bmatrix}. \tag{4}$$

The opposite states are connected by the mirror operator [$\widehat{M}_{xy}(+P) = -P$] in interlayer-sliding ferroelectric materials. Therefore, the in-plane BPVE coefficient is invariant with the change of ferroelectric orders, i.e. $\sigma_{ab}^{c}(+P) = \sigma_{ab}^{c}(-P), a, b, c \in \{x, y\}$, according to Eq. (3). This is strikingly different from the out-of-plane case where the BPVE coefficient is reversed with the switch of ferroelectric orders, as indicated by $\sigma_{bc}^{\prime z}(+P) = -\sigma_{bc}^{\prime z}(-P)$ ( $b, c \in \{x, y\}$ ). The calculated BPVE characters in **Fig. 1** (a) and (c) by a 1D effective model are also consistent with our symmetry analysis (see Supplementary Note 2).

In contrast, for the ferroelectric states correlated by inversion symmetry ($\hat{I}(+P) = -P$), the inversion operator $\hat{I}$ will lead to the reverse of all the vectors (**Fig. 1**(d)), i.e. $\sigma_{bc}^{a}(+P) = -\sigma_{bc}^{a}(-P)$ ($a, b, c \in \{x, y, z\}$) according to Eq. (3). It is the reason why the BPVE coefficients reverse with ferroelectric orders in the conventional ferroelectric materials.

The in-plane BPVE coefficients can be calculated by the standard second-order Kubo formalism theory[32,33], and the out-of-plane BPVE coefficients can be obtained by a modified second-order Kubo formalism theory[31] that was proposed recently. In the following part, we choose a representative material for each case: bilayer MoS$_2$[22,23] (Case 2a) and bilayer-WTe$_2$[14-17] (Case 1b) to perform the numerical first-principles calculations (see Methods section for calculation details).

### First-principles calculation results

**Bilayer MoS$_2$**  Bilayer MoS$_2$ (Case 2a) that is consisted of two identical non-ferroelectric monolayers shows out-of-plane ferroelectricity[13], which has been confirmed in experiments recently[22,23]. Noting that conventional bilayer MoS$_2$ is stacked in A/B form (Case 2b) with inversion symmetry, which is different from the crystal structure discussed here. As shown in **Fig. 2**(a), the out-of-plane ferroelectricity is reversed by the in-plane interlayer motion along the armchair direction. The monolayer possesses the $D_{3h}$ symmetry. However, the interlayer-sliding bilayer breaks the $\widehat{M}_{xy}$ and in-plane $\hat{C}_2$ rotation symmetries, which results in the bilayer

MoS$_2$ only having $C_{3v}$ symmetry ($D_{3h} = C_{3v} + \widehat{M}_{xy} C_{3v}$). Therefore, +P and –P states only have out-of-plane polarization. Moreover, the two ferroelectric states are connected by the $\widehat{M}_{xy}$ symmetry[22,23], which is consistent with the above symmetry analysis. The calculated band structures of the +P and –P states are shown in **Fig. 2** (b), which are invariant with the ferroelectric orders.

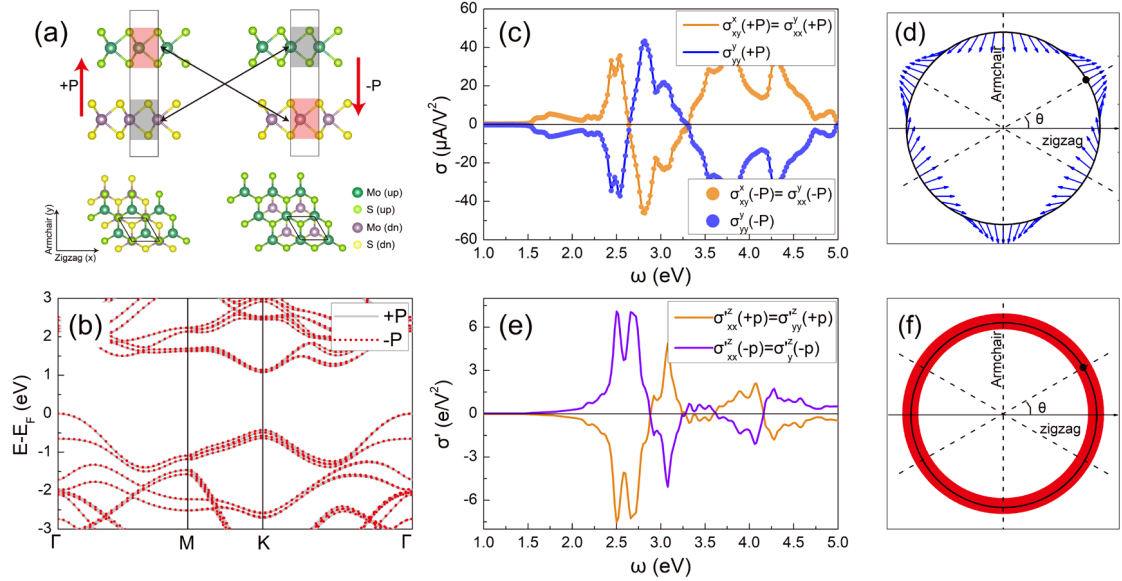

**Fig. 2 Calculated results of ferroelectric bilayer MoS$_2$.** (a) Crystal structures of ferroelectric states (upper panel: side view, lower panel: top view). The arrows denote the transformation of the +P and –P states by interlayer $\widehat{M}_{xy}$ symmetry. The definitions of +P and –P states are adopted from Ref. 22. (b) Band structure of +P and –P states. (c) In-plane and (e) out-of-plane BPVE coefficients. (d) In-plane BPVE photocurrent and (f) out-of-plane BPVE polarization (in arbitrary unit) with the direction of normal-incidence polarization of light in the +P states. $\theta$ in (d) and (f) is the azimuthal angle of the polarization of light relative to the *x* axis, and the dash lines mean the vertical mirror planes.

As shown in **Fig. 2**(c), the calculated in-plane BPVE coefficients with +P and –P states are equal. The in-plane BPVE is invariant with the change of the ferroelectric order as predicted. In addition, $\sigma_{xy}^{x} = \sigma_{xx}^{y} = -\sigma_{yy}^{y}$, and $\sigma_{xx}^{x} = \sigma_{yy}^{x} = \sigma_{xy}^{y} = 0$ for each ferroelectric state due to the C$_{3v}$ symmetry [see Supplementary Note 3]. It is noted that this unswtichable in-plane BPVE character of interlayer-sliding ferroelectric materials is different from those of conventional ferroelectric materials[6-12,34] and antiferromagnets[35,36] where the BPVE is switched with the ferroelectric/ferromagnetic order. The in-plane BPVE photocurrent as a function of the polarization direction of light (see Supplementary Note 4 for details) is shown in **Fig. 2**(d). The in-plane BPVE photocurrent shows the $C_{3v}$ symmetry, and its magnitude is independent of the direction of light.

The calculated out-of-plane BPVE coefficients are shown in **Fig. 2**(e). For each ferroelectric state, $\sigma'^z_{xx} = \sigma'^z_{yy}$, and $\sigma'^z_{xy} = \sigma'^z_{yx} = 0$. Moreover, the out-of-plane BPVE coefficients switch with the ferroelectric order following $\sigma'^z_{xx}(+P) = \sigma'^z_{yy}(+P) = -\sigma'^z_{xx}(-P) = -\sigma'^z_{yy}(-P)$, which is consistent with our symmetry analysis. The out-of-plane BPVE is isotropic, and photo-induced polarization is independent of the polarization direction of light due to $C_{3v}$ symmetry, as shown in **Fig. 2**(f).

Similarly, experimentally prepared bilayer BN[18,19], InSe[20,21], GaSe, *etc.*, possess similar interlayer-sliding ferroelectricity (Case 2a) and the same symmetry. Therefore, similar in-plane and out-of-plane BPVE features are also expected to exhibit in these materials.

**Bilayer WTe₂** Bilayer WTe$_2$ is composed of two 180°-rotated two monolayers (Case 1b), as shown in **Fig. 3(a)**. Even though each monolayer (with $C_{2h} = \{E, \hat{I}, \hat{C}_{2x}, \widehat{M}_{yz}\}$ symmetry) has the inversion $\hat{I}$ symmetry, the bilayer breaks this symmetry. Besides, the interlayer-sliding vector along $b$ axis further breaks $\widehat{M}_{xy}$ and in-plane $\hat{C}_2$ symmetry. Thereby the bilayer has the $C_{1v} = \{E, \widehat{M}_{yz}\}$ symmetry ($C_{2h} = C_{1v} + \widehat{M}_{xy} C_{1v}$).

Interlayer sliding[37,38] induced out-of-plane ferroelectricity in bilayer WTe$_2$ has been verified experimentally[14-17]. The –P state is the mirror image of +P state, as shown in **Fig. 3**(a). **Figure 3**(b) shows the band structures of +P and –P states, where the valence and conduction bands overlap in our calculation.

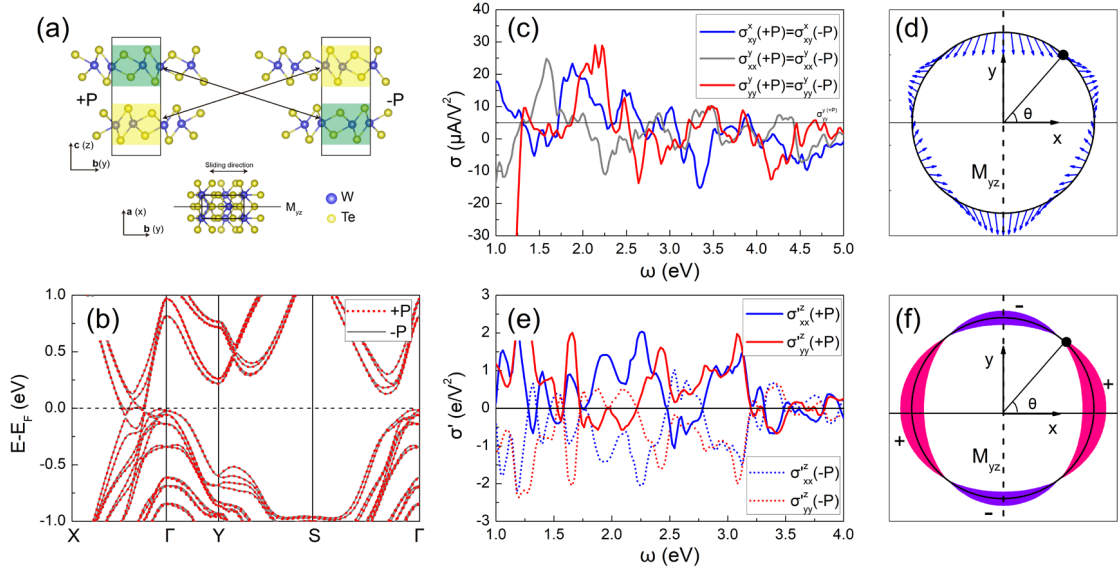

**Fig. 3 Calculated results of ferroelectric bilayer WTe$_2$.** (a) Crystal structure. +P and –P states are adopted from Ref. 37. Rectangle in (a) represents the unit cell, and the double-arrows line means the monolayer transformation under $\widehat{M}_{xy}$ symmetry. The lower panel in (a) denotes the top view of bilayer WTe$_2$ with the +P state. (b) Band structures of +P and –P states. (c) In-plane and (e) out-of-plane BPVE coefficients. (d) In-plane BPVE photo-current and (f) out-of-plane BPVE polarization of the +P states with the different direction of polarization of light. The length of arrows in (d) indicates the magnitude of in-plane BPVE response (in arbitrary unit), where $\sigma_{xy}^{x}:\sigma_{xx}^{y}:\sigma_{yy}^{y}=1:0.7:-1.6$. The width and the color of the ribbon in (f) denote the magnitude and direction of out-of-plane BPVE response, and $\sigma_{xx}^{\prime z}:\sigma_{yy}^{\prime z}=1:-0.6$.

The calculated in-plane BPVE coefficients are shown in **Fig. 3** (c). There are three independent tensor elements $\sigma_{xy}^{x}$, $\sigma_{xx}^{y}$, $\sigma_{yy}^{y}$, and $\sigma_{xx}^{x}=\sigma_{yy}^{x}=\sigma_{xy}^{y}=0$ due to the C$_{1v}$ symmetry. The calculated in-plane BPVE coefficients are the same in two opposite ferroelectric orders: $\sigma_{xy}^{x}(+P)=\sigma_{xy}^{x}(-P)$, $\sigma_{xx}^{y}(+P)=\sigma_{xx}^{y}(-P)$, $\sigma_{yy}^{y}(+P)=\sigma_{yy}^{y}(-P)$, which is consistent with the above symmetry analysis. The in-plane BPVE photo-current with the direction of polarization of light is shown in **Fig. 3** (d), which shows $\widehat{M}_{yz}$ symmetry, and its magnitude is dependent on the light (see Supplementary Note 4 for details).

On the contrary, the out-of-plane BPVE coefficients switch with the change of ferroelectric order, $\sigma_{xx}^{\prime z}(+P)=-\sigma_{xx}^{\prime z}(-P)$, $\sigma_{yy}^{\prime z}(+P)=-\sigma_{yy}^{\prime z}(-P)$ [**Fig. 3** (e)]. Unlike the bilayer MoS$_2$, the out-of-plane BPVE coefficients show anisotropy $\sigma_{xx}^{\prime z}\neq\sigma_{yy}^{\prime z}$ ($\sigma_{xy}^{\prime z}=\sigma_{yx}^{\prime z}=0$) for each ferroelectric state due to the absence of in-plane C$_3$ symmetry, leading to the out-of-plane BPVE polarization is dependent on the polarization of light as shown in **Fig. 3** (f).

Recently, $ZrI_2$, a sister material of polar semimetals $WTe_2$, is also demonstrated[39-41] to have interlayer-sliding ferroelectricity. Similar BPVE behaviors with bilayer $WTe_2$ are expected to be observed due to the same symmetry and interlayer-stacking way (Case 1b). The calculated values of the in-plane BVPE coefficient of bilayer $MoS_2$ and bilayer $WTe_2$ are comparable to the reported monolayer SnTe[42,43] and 3D conventional ferroelectric materials (such as $BiFeO_3$[44], $BiTiO_3$, and $PbTiO_3$[2]), and out-of-plane coefficients are comparable to twisted bilayers graphene[31], which are detectable experimentally.

## Discussion and conclusion

According to the above two specific examples, the in-plane BPVE is invariant, while the out-of-plane BPVE reverses with the change of ferroelectric order in interlayer-sliding ferroelectric materials. The study of the ferroelectric order and the BPVE in mirror-symmetry-related ferroelectric materials allows the comprehension of the relationship between them in another view.

Moreover, these BPVE characters in 2D interlayer-siding ferroelectric materials will lead to distinctive photovoltaic phenomena and applications in photoelectric devices. The in-plane BPVE photocurrents in a single crystal of interlayer-siding ferroelectric materials are invariant with two opposite ferroelectric domains, thus the photovoltaic currents between different domains superimpose rather than cancel, as shown in **Fig. 4**. Since ferroelectric domains are often unavoidable in real materials, especially for the interlayer-sliding ferroelectric materials fabricated via the "tear and stack" method[18,19,23,45]. Therefore, this unswitchable BPVE feature boosts the development of large-scale photoelectric devices that are immune to the polarization of ferroelectric domains. Nevertheless, the out-of-plane BPVE inverses between the opposite ferroelectric orders (**Fig. 4**), which can be utilized to construct switchable photoelectric devices.

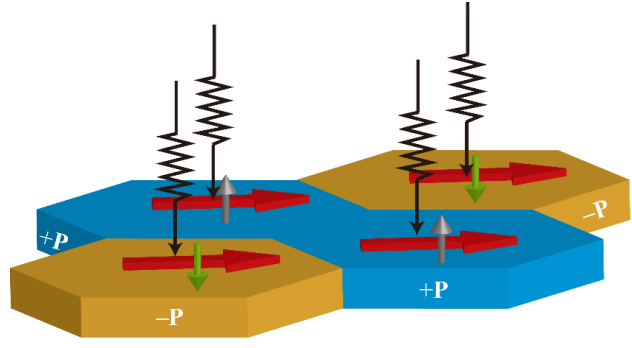

**Fig. 4 In-plane and out-of-plane BPVE response with opposite ferroelectric domains for interlayer-sliding ferroelectric materials.** The horizontal and vertical arrows denote the in-plane BPVE current and out-of-plane BPVE polarization, respectively.

In conclusion, we use the abstract bilayer crystal model to analyze the symmetries of the four most common cases of bilayer vdW materials with interlayer sliding, and found two cases (Case 1b and Case 2a) can have the possibility to achieve the interlayer sliding ferroelectricity, which the interlayer sliding ferroelectric materials found in experiments fall into. We revealed that the opposite ferroelectric states in the interlayer-sliding ferroelectric materials are linked by the horizontal mirror symmetry, leading to the in-plane BPVE being invariant while the out-of-plane BPVE switches with opposite ferroelectric states. Our theoretical study provides a strategy towards a comprehensive understanding of the relationship between the ferroelectric order and BPVE in 2D interlayer sliding ferroelectric materials. Moreover, the switchable out-of-plane BPVE can be used to design switchable photoelectric devices. On the other hand, the in-plane BPVE is robust against the ferroelectric order, and the unswitchable character is beneficial to construct larger-scale photoelectric devices.

## Methods

### In-plane and out-of-plane BPVE theory

According to the second-order Kubo formalism theory[32,33], $\sigma_{bc}^{a}$ in the Eq. (1) of the main text can be expressed as

$$\sigma_{bc}^{a} \equiv \frac{1}{2}\text{Re}\left\{\chi_{bc}^{a} + \chi_{cb}^{a}\right\}, \qquad (5)$$

where

$$\chi_{bc}^{a}(\omega) = -\frac{e^3}{\omega^2}\int\frac{d\mathbf{k}}{(2\pi)^3}\sum_{mnl}\frac{f_{lm}v_{lm}^{b}}{E_{ml}-\hbar\omega+i\delta}\left(\frac{v_{mn}^{a}v_{nl}^{c}}{E_{mn}+i\delta}-\frac{v_{mn}^{c}v_{nl}^{a}}{E_{nl}+i\delta}\right), \qquad (6)$$

$\delta = \hbar/\tau$, and $\tau$ is the lifetime, $\omega$ is the frequency of the light, $f_{lm}$ and $E_{ml}$ are the difference of occupation number and band energy between bands $l$ and $m$.

Within a modified second-order Kubo formalism theory[31], the out-of-plane BPVE coefficient in Eq. (2) of the main text can be expressed as

$$\sigma_{bc}^{\prime z} \equiv \frac{1}{2}\text{Re}\left\{\chi_{bc}^{\prime z} + \chi_{cb}^{\prime z}\right\}, \qquad (7)$$

and

$$\chi_{bc}^{\prime z} = \frac{e^2}{\omega^2} \int \frac{d\mathbf{k}}{(2\pi)^3} \times \sum_{imnl} \frac{f_{lm} v_{lm}^b}{E_{ml} - \hbar\omega + i\delta} \left( \frac{(\hat{p}_{i,z})_{nm} v_{nl}^c}{E_{mn} + i\delta} - \frac{v_{mn}^c (\hat{p}_{i,z})_{nm}}{E_{nl} + i\delta} \right), \qquad (8)$$

where $\hat{p}_{i,z} = er_{i,z}$ is a position operator, and $r_{i,z}$ is the z-component position of the i-th atom. We define the middle of the 2D materials as the origin: $\sum_i r_{i,z} = 0$. Eq. (6) is very similar to Eq. (8), except for the position operator $\hat{p}_{i,z}$ instead of the velocity operator.

**First-principles calculations**

The first-principles calculations based on density functional theory (DFT) are performed by using the VASP package. General gradient approximation (GGA) according to the Perdew-Burke-Ernzerhof (PBE) functional is used. The energy cutoff of the plane wave basis is set to 400 eV. The Brillouin zone is sampled with a 12×12×1 (12×8×1) mesh of **k**-points for MoS$_2$ (WTe$_2$). To simulate the monolayers, vacuum layers (~15 Å) are introduced. The vdW force with DFT-D2 correction is considered. Spin-orbital coupling effects are considered in our calculation.

The DFT Bloch wave functions are iteratively transformed into maximally localized Wannier functions by the Wannier90 code[46,47]. Mo-*d* and S-*p* (W-*d* and Te-*p*) orbitals are used to construct the Wannier functions for MoS$_2$ (WTe$_2$). The in-plane and out-of-plane BPVE coefficients are calculated by our own program WNLOP (Wannier Non-Linear Optics Package) based on effective tight-binding (TB) Hamiltonian. Convergence test of k-mesh is performed, and 500×500×1 (600×300×1) k-mesh is sufficient in BPVE calculations of MoS$_2$ (WTe$_2$). We adopt δ=0.02 eV for Eq. (6) and Eq. (8) in our calculation to take into account various relaxation processes.

The 3D-like BPVE coefficients are obtained assuming an active single-layer with the thickness of $L_{active}$[13]

$$\sigma_{3D} = \frac{L_{slab}}{L_{active}} \sigma_{slab} \tag{9}$$

where $\sigma_{slab}$ is the calculated BPVE coefficient, and $L_{slab}$ ($L_{active} < L_{slab}$) is the effective thickness slab.

## Acknowledgments

The authors acknowledge the High-performance Computing Platform of Anhui University for providing computing resources. We thank Bin Xu, Ding-Fu Shao, and Yuan-Jun Jin for their useful discussions. This work is supported by the National Basic Research Program of China under No. 11947212, and in part by the Joint Funds of the National Natural Science Foundation of China and the Chinese Academy of Sciences Large-Scale Scientific Facility under Grant No. U1932156, and in part by the Natural Science Foundation of Anhui Province under Grant No. 2008085QA29. Y. G. and R.C.X. acknowledge the startup foundation from USTC and AHU, respectively.

## Author contributions

R.C.X. conceived the project and edited the code and performed the first-principles calculations. R.C.X. performed the symmetry analysis, and Y.G. checked it. R.C.X., Y.G., and H. L. discussed the results and the writing. The manuscript was written through the contributions of all authors. All authors have approved the final version of the manuscript.

## Additional information

**Supplementary information**   The online version contains supplementary material available at https:/xxxxx.

# Supplementary Information

## Supplementary Note 1: Symmetry analysis of bilayer materials with interlayer sliding using *abstract bilayer crystal model*

As stated in the main text, the non-ferroelectric monolayer materials always belong to the following two cases:

1. Monolayer has inversion symmetry but no horizontal mirror symmetry, including transition-metal dichalcogenides and dihalides in $1T$ phase [short for $1T$-$MX_2$, such as $CdI_2$ and $PtS_2$, **Fig. S1**(a)], transition-metal dichalcogenides in $1T'$ phase [short for $1T'$-$MX_2$, such as $MoTe_2$, $WTe_2$, and $ZrI_2$, Fig. S1(b)], and metal trihalide [short for $MX_3$, such as $BiI_3$ and $CrI_3$, **Fig. S1**(c)];

2. Monolayer has horizontal mirror symmetry but no inversion symmetry, including BN [**Fig. S2**(a)], transition-metal dichalcogenides in $1H$ phase [short for $1H$-$MX_2$, such as $MoS_2$, $MoSe_2$, $WS_2$ and $WSe_2$, **Fig. S2**(b)], and post-transition metal chalcogenides [short for MX, such InSe, GaSe and GaS, **Fig. S2**(c)].

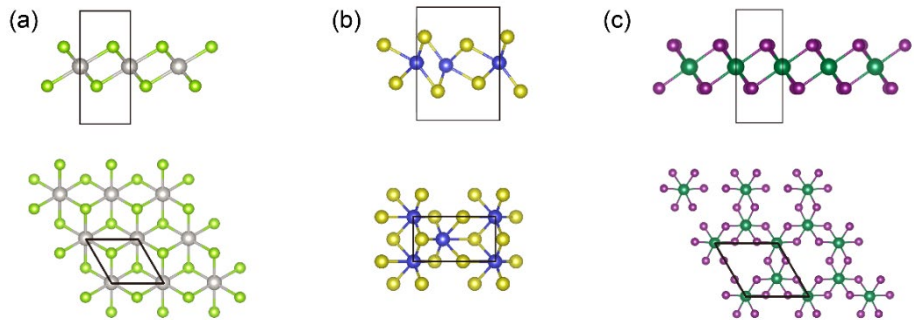

Fig. S1 Crystal structure of monolayer of (a) $1T$-$MX_2$, (b)$1T'$-$MX_2$, and (c) $MX_3$

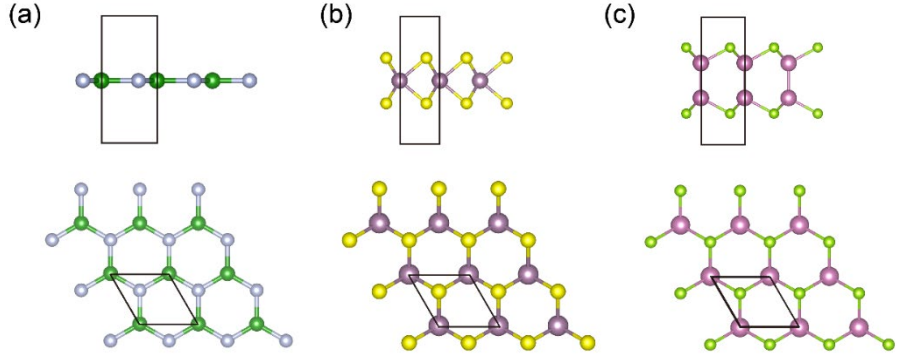

Fig. S2 Crystal structure of monolayer of (a) BN, (b) 1$H$-MX$_2$, and (c) MX

In each case, the stacking way of the bilayer can be further divided into two subcases (A/A or A/B, where A or B is the crystal structure, and $B = \hat{C}_{2z} A$).

The bilayer materials with interlayer sliding can be described with an *abstract bilayer crystal model* (as shown in **Fig. S3**) by:

$$[(A, +\mathbf{r}_{//}), (B, -\mathbf{r}_{//})], \quad \text{S(10)}$$

where the parentheses $(A, +\mathbf{r}_{//})$ and $(B, -\mathbf{r}_{//})$ represent the states of the first (top) and second (button) layers respectively, where the A or B represents the crystal structure of the monolayer and $\mathbf{r}_{//}$ represents the in-plane sliding vector.

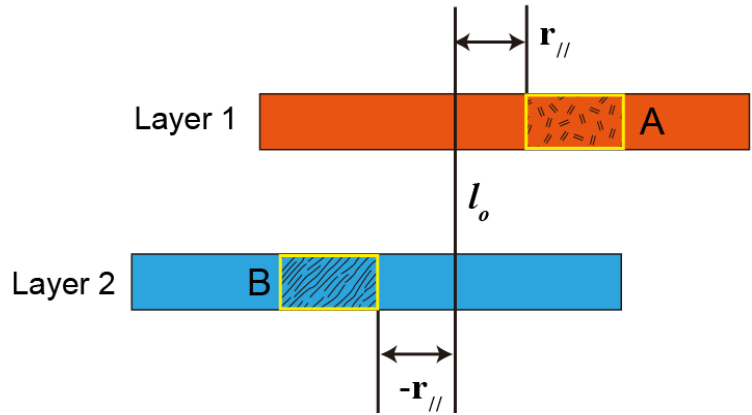

Fig. S3  Schematic of abstract bilayer crystal model. The yellow rectangles denote the unit cells, and line $l_o$ is the original line.

In the interlayer-sliding motion, atoms in each monolayer do not move relative to themselves, so the symmetry of each layer does not change. The interlayer-sliding vector in Eq. S(1) is a random number that does not equal the integral or half-integral lattice vector, so the in-plane C₂ symmetry is broken in the bilayer system. In the following, we analyze the symmetries of the bilayer materials with interlayer sliding, specifically the inversion $\hat{I}$ and mirror $\widehat{M}_{xy}$ symmetry. In the following, we analyze the symmetry of the four cases in Table I the main text with the abstract bilayer crystal model.

**Case 1: Monolayer with inversion symmetry without M$_{xy}$ symmetry**

**Case 1a: A/A stacking**

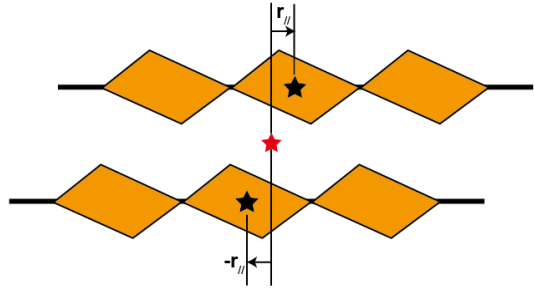

Fig. S4  2D bilayer system stacking by Case 1a. The quadrilateral means abstract unit cell of the monolayer, and star means the inversion symmetry.

In this case, the bilayer materials are composed of two identical monolayers which both have the inversion symmetry, as shown in **Fig. S4**. According to the definition of Eq. S(1), the bilayer crystal structure can be written as

$$S_+ : [(A, +\mathbf{r}_{//}), (A, -\mathbf{r}_{//})]. \qquad \text{S(11)}$$

Similarly, the bilayer with the opposite interlayer-sliding direction can be described as

$$S_- : [(A, -\mathbf{r}_{//}), (A, +\mathbf{r}_{//})]. \qquad \text{S(12)}$$

Now, we perform inversion operation $\hat{I}$ (in the middle of the bilayer and at the original axis) to the bilayer with S$_+$ state,

$$\begin{aligned}
\hat{I} S_+ &= \hat{I}[(A, +\mathbf{r}_{//}), (A, -\mathbf{r}_{//})] \\
&= [\hat{I}(A, -\mathbf{r}_{//}), \hat{I}(A, +\mathbf{r}_{//})] \\
&= \left[\left(\hat{I}A, \hat{I}(-\mathbf{r}_{//})\right), \left(\hat{I}A, \hat{I}(+\mathbf{r}_{//})\right)\right] \qquad \text{S(13)} \\
&= [(A, +\mathbf{r}_{//}), (A, -\mathbf{r}_{//})] \\
&= S_+.
\end{aligned}$$

The formula at the second line runs because the inversion $\hat{I}$ operator between the two layers switches the first layer and the second layer. The formula at the fourth line holds because each monolayer is invariant under the inversion operator: $\hat{I}A = A$, and the sliding vector $\mathbf{r}_{//}$ reverse under the inversion operator. Therefore, the $S_+$ state is invariant under the inversion symmetry.

Now, we perform mirror operation $\widehat{M}_{xy}$ to $S_+$ state:

$$\begin{aligned}\widehat{M}_{xy}S_+ &= \widehat{M}_{xy}[(A,+\mathbf{r}_{//}),(A,-\mathbf{r}_{//})]\\ &=[\widehat{M}_{xy}(A,-\mathbf{r}_{//}),\widehat{M}_{xy}(A,+\mathbf{r}_{//})]\\ &=\left[\left(\widehat{M}_{xy}A,\widehat{M}_{xy}(-\mathbf{r}_{//})\right),\left(\widehat{M}_{xy}A,\widehat{M}_{xy}(+\mathbf{r}_{//})\right)\right]\\ &=[(B,-\mathbf{r}_{//}),(B,+\mathbf{r}_{//})].\end{aligned} \quad \text{S(14)}$$

Because the monolayers have no $\widehat{M}_{xy}$ symmetry, the monolayer crystal structure changes from A to B, therefore the bilayer also has no $\widehat{M}_{xy}$ symmetry.

In all, under the inversion operator $\hat{I}$, $S_+$ changes into itself whatever $\mathbf{r}_{//}$, so this kind of bilayer has inversion symmetry and cannot process ferroelectricity.

**Case 1b: A/B stacking**

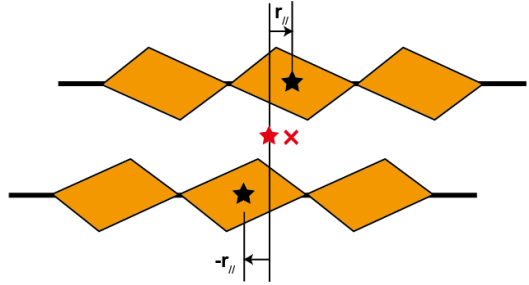

Fig. S5   2D bilayer material stacking by Case 1b.

In this case, each monolayer is related to each other by $\widehat{C}_{2z}$ operator, as shown in **Fig. S5** (A/B stacking). The two opposite sliding states can be described as

$$\begin{cases} S_+:[(A,+\mathbf{r}_{//}),(B,-\mathbf{r}_{//})],\\ S_-:[(A,-\mathbf{r}_{//}),(B,+\mathbf{r}_{//})],\end{cases} \quad \text{S(15)}$$

where

$$\begin{cases} \hat{I}A = A, \\ \hat{I}B = B, \end{cases} \quad \text{S(16)}$$

and each monolayer crystal structure is switched by $\widehat{C}_{2z}$ symmetry:

$$\begin{cases} \widehat{C}_{2z}A = B, \\ \widehat{C}_{2z}B = A. \end{cases} \quad \text{S(17)}$$

For a system with inversion symmetry $\hat{I}$, $\widehat{M}_{xy}$ and $\widehat{C}_{2z}$ operators are equivalent, because $\hat{I}\widehat{C}_{2z}=\widehat{M}_{xy}$. Therefore, the monolayer crystal structure is also switched under the $\widehat{M}_{xy}$ operator:

$$\begin{cases} \widehat{M}_{xy}A = B, \\ \widehat{M}_{xy}B = A. \end{cases} \quad \text{S(18)}$$

Now, we perform inversion operation $\hat{I}$ to the $S_+$ state,

$$\begin{aligned} \hat{I}S_+ &= \hat{I}[(A,+\mathbf{r}_{//}),(B,-\mathbf{r}_{//})] \\ &= \left[\hat{I}(B,-\mathbf{r}_{//}), \hat{I}(A,+\mathbf{r}_{//})\right] \\ &= [(B,+\mathbf{r}_{//}),(A,-\mathbf{r}_{//})]. \end{aligned} \quad \text{S(19)}$$

$\hat{I}S_+ \neq S_+$, so bilayer $S_+$ does not have the inversion symmetry whatever $\mathbf{r}_{//}$. Besides, $\hat{I}S_+ \neq S_-$, therefore the two opposite interlayer-sliding states are not connected by inversion operation.

Performing $\hat{M}_{xy}$ operation to the $S_+$ state,

$$\begin{aligned} \widehat{M}_{xy}S_+ &= \widehat{M}_{xy}[(A,+\mathbf{r}_{//}),(B,-\mathbf{r}_{//})] \\ &= \left[\widehat{M}_{xy}(B,-\mathbf{r}_{//}), \widehat{M}_{xy}(A,+\mathbf{r}_{//})\right] \\ &= [(A,-\mathbf{r}_{//}),(B,+\mathbf{r}_{//})] \\ &= S_-. \end{aligned} \quad \text{S(20)}$$

The formula at the second line holds because the $\hat{M}_{xy}$ operator between the two layers will switch the first layer and the second layer. The formula after the third line runs because the monolayer crystal types switch under $\hat{M}_{xy}$ [Eq. S(9)], and $\mathbf{r}_{//}$ is invariant under $\hat{M}_{xy}$. Therefore, under the $\hat{M}_{xy}$ operation, S+ cannot transform into itself but convert to the S- state. Besides, the bilayer has the $\hat{M}_{xy}$ symmetry if $\mathbf{r}_{//} = 0$.

To sum up, the bilayer crystal structure in Case 1b does not have both inversion and $\hat{M}_{xy}$ symmetry, but the two opposite interlayer-sliding states are connected by $\hat{M}_{xy}$ symmetry.

**Case 2 Monolayer without inversion symmetry but has the M$_{xy}$ symmetry**

**Case 2a: A/A stacking**

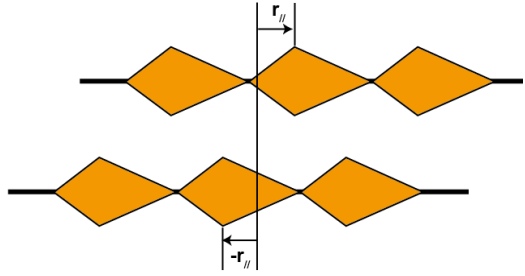

Fig. S6  2D bilayer material stacking by Case 2a.

In this case, the bilayer material is composed of two identical monolayers without inversion symmetry but with the $\hat{M}_{xy}$ symmetry, as shown in **Fig. S6**. The two opposite interlayer-sliding states of this case can be described as

$$\begin{cases} S_+ : [(A,+\mathbf{r}_{//}),(A,-\mathbf{r}_{//})], \\ S_- : [(A,-\mathbf{r}_{//}),(A,+\mathbf{r}_{//})]. \end{cases} \quad \text{S(21)}$$

The monolayer crystal structure is invariant under the $\hat{M}_{xy}$ operation: $\widehat{M}_{xy}A = A$.

Performing $\hat{I}$ operation to the S+ state, we will get

$$\begin{aligned} \hat{I}S_+ &= \hat{I}[(A,+\mathbf{r}_{//}),(A,-\mathbf{r}_{//})] \\ &= \left[ \hat{I}(A,-\mathbf{r}_{//}), \hat{I}(A,+\mathbf{r}_{//}) \right] \\ &= [(B,+\mathbf{r}_{//}),(B,-\mathbf{r}_{//})], \end{aligned} \quad \text{S(22)}$$

where the monolayer has no inversion symmetry, and A transforms into B under $\hat{I}$. We find $\hat{I}S_+ \neq S_+$, and $\hat{I}S_+ \neq S_-$. Therefore, S+ does not have the inversion symmetry,

S6

and the two opposite interlayer-sliding states are not related to each other by the inversion operation either.

We perform the $\widehat{M}_{xy}$ operator to $S_+$ state,

$$\begin{aligned}\widehat{M}_{xy}S_+ &= \widehat{M}_{xy}[(A,+\mathbf{r}_{//}),(A,-\mathbf{r}_{//})] \\ &= \left[\widehat{M}_{xy}(A,-\mathbf{r}_{//}),\widehat{M}_{xy}(A,+\mathbf{r}_{//})\right] \\ &= [(A,-\mathbf{r}_{//}),(A,+\mathbf{r}_{//})] \\ &= S_-.\end{aligned} \quad \text{S(23)}$$

Therefore, $S_+$ has no $\widehat{M}_{xy}$ symmetry if $\mathbf{r}_{//} \neq 0$, but $S_+$ can convert to $S_-$ by $\widehat{M}_{xy}$.

In summary, the bilayer 2D materials in Case 2a do not have both inversion $\hat{I}$ and $\hat{M}_{xy}$ symmetry, but the two opposite interlayer-sliding states are connected by $\hat{M}_{xy}$ operation.

**Case 2b A/B stacking**

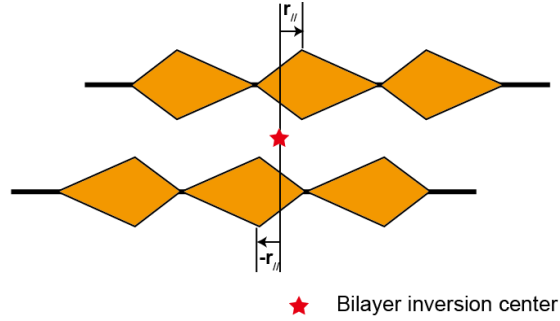

Fig. S7    2D bilayer material stacking by Case 2b

In this case, the bilayer is constituted by two 180°-rotated monolayers (A/B stacking), as shown in **Fig. S7**. The two opposite interlayer-sliding states of this case can be described as

$$\begin{cases} S_+ : [(A,+\mathbf{r}_{//}),(B,-\mathbf{r}_{//})], \\ S_- : [(A,-\mathbf{r}_{//}),(B,+\mathbf{r}_{//})], \end{cases} \quad \text{S(24)}$$

where each layer has the $\widehat{M}_{xy}$ symmetry:

$$\begin{cases} \widehat{M}_{xy} A = A, \\ \widehat{M}_{xy} B = B. \end{cases} \quad \text{S(25)}$$

and each monolayer is switched by $\widehat{C}_{2z}$ symmetry

$$\begin{cases} \widehat{C}_{2z}A = B, \\ \widehat{C}_{2z}B = A. \end{cases} \quad \text{S(26)}$$

In a system with $\widehat{M}_{xy}$ symmetry, $\widehat{C}_{2z}$ and $\hat{I}$ are equivalent, because $\widehat{M}_{xy}\widehat{C}_{2z} = \hat{I}$. Therefore, the monolayer crystal types are also switched under $\hat{I}$:

$$\begin{cases} \hat{I}A = B, \\ \hat{I}B = A. \end{cases} \quad \text{S(27)}$$

Performing $\hat{I}$ operation to the S$_+$ state,

$$\begin{aligned}
\hat{I}S_+ &= \hat{I}[(A, +\mathbf{r}_{//}), (B, -\mathbf{r}_{//})] \\
&= \left[\hat{I}(B, -\mathbf{r}_{//}), \hat{I}(A, +\mathbf{r}_{//})\right] \\
&= [(A, +\mathbf{r}_{//}), (B, -\mathbf{r}_{//})] \\
&= S_+.
\end{aligned} \quad \text{S(28)}$$

S$_+$ turns itself into itself under inversion operation whatever $\mathbf{r}_{//}$, so this structure has inversion symmetry, even though the monolayer has no inversion symmetry. Therefore, the bilayer in Case 2b cannot have ferroelectricity and BPVE.

Besides, we perform inversion operation $\widehat{M}_{xy}$ to S$_+$ state:

$$\begin{aligned}
\widehat{M}_{xy}S_+ &= \widehat{M}_{xy}[(A, +\mathbf{r}_{//}), (B, -\mathbf{r}_{//})] \\
&= [\widehat{M}_{xy}(B, -\mathbf{r}_{//}), \widehat{M}_{xy}(A, +\mathbf{r}_{//})] \\
&= \left[\left(\widehat{M}_{xy}B, \widehat{M}_{xy}(-\mathbf{r}_{//})\right), \left(\widehat{M}_{xy}A, \widehat{M}_{xy}(+\mathbf{r}_{//})\right)\right] \\
&= [(B, -\mathbf{r}_{//}), (A, +\mathbf{r}_{//})] \neq S_+
\end{aligned} \quad \text{S(29)}$$

Even though monolayer lattices A and B have the $\widehat{M}_{xy}$ symmetry, this kind of bilayer has no $\widehat{M}_{xy}$ symmetry.

The above four cases are summarized in Table I of the main text.

Last but not least, in real materials such as bilayer MoS$_2$, BN, InSe, GaSe (Case 2a), except the sliding vector $\mathbf{r}_{//}$, there is also a half lattice vector shift $1/2\mathbf{a}_\perp$ for the top layer in the positive and negative ferroelectric state. $1/2\mathbf{a}_\perp$ is vertical to the interlayer-sliding direction [see **Fig. 2(a)** in the main text]. The two interlayer-sliding states in Eq. S(12) of Case 2a are changed into

$$\begin{cases} S_+ : [(A, +\mathbf{r}_{//} + 1/2\mathbf{a}_\perp), (A, -\mathbf{r}_{//} + 1/2\mathbf{a}_\perp)], \\ S_- : [(A, -\mathbf{r}_{//} + 1/2\mathbf{a}_\perp), (A, +\mathbf{r}_{//} + 1/2\mathbf{a}_\perp)], \end{cases} \qquad S(30)$$

The crystals have the translation symmetry with integer lattice vector, so we can shift the second/first layer with an integral constant vector $-\mathbf{a}_\perp$ for $S_+/S_-$, then

$$\begin{cases} S'_+ : [(A, +\mathbf{r}_{//} + 1/2\mathbf{a}_\perp), (A, -\mathbf{r}_{//} - 1/2\mathbf{a}_\perp)], \\ S'_- : [(A, -\mathbf{r}_{//} - 1/2\mathbf{a}_\perp), (A, +\mathbf{r}_{//} + 1/2\mathbf{a}_\perp)]. \end{cases} \qquad S(31)$$

Now, Eq. S(22) here is equal to Eq. S(12) in Case 2a.

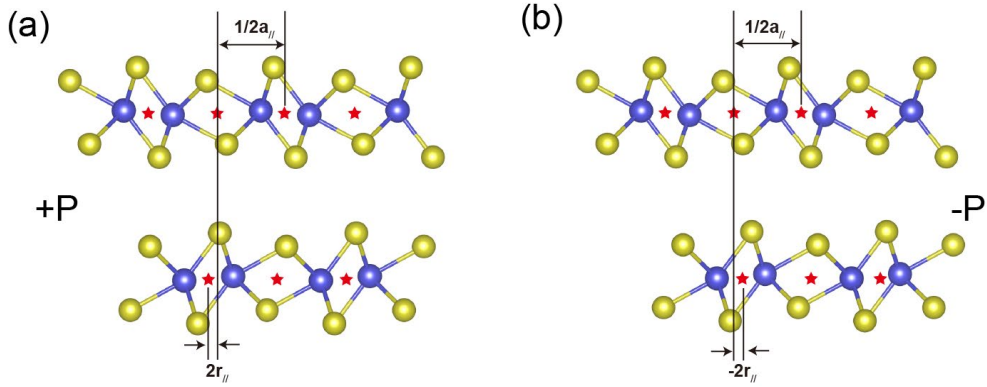

Fig. S8  Interlayer sliding of bilayer WTe$_2$ with (a) +P and (b) –P states, where $2\mathbf{r}_{//}$ means the interlayer sliding vector compared to Fig. S5.

For bilayer WTe$_2$ (Case 1b), except for the half lattice vector shift $1/2\mathbf{a}_\perp$ [See Fig. 3(a) in the main text], there also is a half lattice vector interlayer shift $1/2\mathbf{a}_{//}$ that is parallel to the interlayer-sliding direction [**Fig. S8**]. The two interlayer-sliding states for bilayer WTe$_2$ are

$$\begin{cases} S_+ : [(A, +\mathbf{r}_{//} + 1/2\mathbf{a}_\perp + 1/2\mathbf{a}_{//}), (B, -\mathbf{r}_{//} + 1/2\mathbf{a}_\perp + 1/2\mathbf{a}_{//})], \\ S_- : [(A, -\mathbf{r}_{//} + 1/2\mathbf{a}_\perp + 1/2\mathbf{a}_{//}), (B, +\mathbf{r}_{//} + 1/2\mathbf{a}_\perp + 1/2\mathbf{a}_{//})], \end{cases} \qquad S(32)$$

Similarly, if we shift the second/first layer with $-\mathbf{a}_\perp - \mathbf{a}_{//}$ for $S_+/S_-$, the two new interlayer-sliding states are

$$\begin{cases} S'_+ : [(A, +\mathbf{r}_{//} + 1/2\mathbf{a}_\perp), (B, -\mathbf{r}_{//} - 1/2\mathbf{a}_\perp - 1/2\mathbf{a}_{//})], \\ S'_- : [(A, -\mathbf{r}_{//} - 1/2\mathbf{a}_\perp - 1/2\mathbf{a}_{//}), (B, +\mathbf{r}_{//} + 1/2\mathbf{a}_\perp)]. \end{cases} \qquad S(33)$$

Here, Eq. S(24) is equal to Eq. S(6) in Case 1b. Therefore, the above rule of Case 1b is also suitable to the bilayer WTe$_2$.

## Supplementary Note 2: Effect model for 1D interlayer sliding

The 1D single chain in **Fig. 1**(a) of the main text is very similar to the SSH (Su-Schrieffer-Heeger) Hamiltonian model

$$\hat{H} = \varepsilon_a \sum_i \hat{a}_i^\dagger \hat{a}_i + \varepsilon_b \sum_i \hat{b}_i^\dagger \hat{b}_i + \sum_i \left( J\hat{a}_i^\dagger \hat{b}_i + J'\hat{a}_i^\dagger \hat{b}_{i-1} + h.c. \right), \quad S(34)$$

i.e.

$$H_0 = \begin{bmatrix} \varepsilon_a & T_{ab} \\ T_{ba} & \varepsilon_b \end{bmatrix}, \quad S(35)$$

where $T_{ab} = T_{ba}^* = J\exp(ik_x) + J'\exp(-ik_x)$, J and J' mean the nearest atom hopping, and $\varepsilon_a$ and $\varepsilon_b$ mean the onsite energies for $a$ (anion) and $b$ (cations) atoms.

For the bi-chain with +P which considering the nearest inter-chain hopping in Fig. 1(a) of the main text, the corresponding effect model is

$$H(+P) = \begin{bmatrix} \varepsilon_a & T_{ab} & 0 & 0 \\ T_{ba} & \varepsilon_b & t_\perp & 0 \\ 0 & t_\perp & \varepsilon_a & T_{ab} \\ 0 & 0 & T_{ba} & \varepsilon_b \end{bmatrix}, \quad S(36)$$

where the basis is $\{\varphi_a^{up}, \varphi_b^{up}, \varphi_a^{dn}, \varphi_b^{dn}\}$, and $t_\perp$ is the nearest hopping parameter between two chains as shown in **Fig. S9**(a). The effective model in bi-chain of –P in **Fig. 1**(c) of the main text is

$$H(-P) = \begin{bmatrix} \varepsilon_a & T_{ab} & 0 & t_\perp \\ T_{ba} & \varepsilon_b & 0 & 0 \\ 0 & 0 & \varepsilon_a & T_{ab} \\ t_\perp & 0 & T_{ba} & \varepsilon_b \end{bmatrix}. \quad S(37)$$

The $H(+P)$ and $H(-P)$ are correlated by the mirror symmetry

$$\widehat{M}_z = \begin{bmatrix} 0 & I_0 \\ I_0 & 0 \end{bmatrix} = \begin{bmatrix} 0 & 0 & 1 & 0 \\ 0 & 0 & 0 & 1 \\ 1 & 0 & 0 & 0 \\ 0 & 1 & 0 & 0 \end{bmatrix} \quad S(38)$$

i.e.

$$\widehat{M}_z H(+P) \widehat{M}_z = H(+P). \tag{S(39)}$$

**Fig. S9**(b) shows the band structure of bi-chain. The in-plane and out-of-plane BPVE coefficients are shown in **Fig. S9**(c) and (d). Consistent with our symmetry analysis, the in-plane BPVE coefficients are invariant while the out-of-plane BPVE coefficients change with the ferroelectric order.

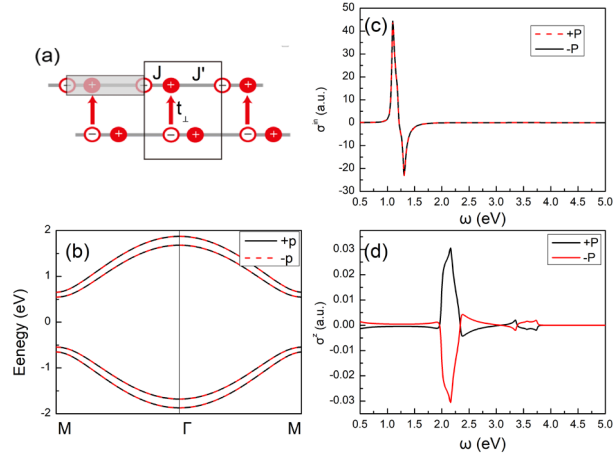

Fig. S9 (a) 1D bi-chain SSH model. (b) Band structure. (c) In-plane and (d) out-of-plane BPVE coefficients. $\varepsilon_a = -\varepsilon_b = 0.5 eV$, $J = 1 eV$, $J' = 0.7 eV$, $t_\perp = 0.2 eV$.

## Supplementary Note 3: Symmetry analysis of the BPVE coefficients

Except for the symmetry requirement shown the Eq. (3) of the main text, the BPVE tensor $\left[\sigma_{bc}^a\right]_{3\times3\times3}$ elements have the subscripts-switch symmetry, *i.e.* $\sigma_{jk}^i = \sigma_{kj}^i$. Therefore, the 3×3×3 $\sigma_{bc}^a$ tensor can be contracted as a 3×6 matrix $\left[\sigma_{jk}^i\right]_{3\times6}$, and the elements in each row are in the sequence of $\sigma_{xx}^i$, $\sigma_{yy}^i$, $\sigma_{zz}^i$, $\sigma_{yz}^i$, $\sigma_{xz}^i$ and $\sigma_{xy}^i$.

Acoordingly, BPVE tensor of bilayer ferroelectric MoS$_2$ with C$_{3v}$ symmetry is

$$\left[\sigma_{jk}^i\right]_{3\times6} = \begin{bmatrix} 0 & 0 & 0 & 0 & \sigma_{xz}^x & -\sigma_{yy}^y \\ -\sigma_{yy}^y & \sigma_{yy}^y & 0 & \sigma_{xz}^x & 0 & 0 \\ \sigma_{xx}^{\prime z} & \sigma_{xx}^{\prime z} & \sigma_{zz}^z & 0 & 0 & 0 \end{bmatrix}. \tag{S(40)}$$

The tensor elements in red mean the in-plane BPVE coefficients, and $\sigma_{xy}^x = \sigma_{xx}^y = -\sigma_{yy}^y$, and $\sigma_{xx}^x = \sigma_{yy}^x = \sigma_{xy}^y = 0$. The elements in blue in Eq. S(31) represent the out-of-plane

BPVE coefficients under the normal incidence of light, and we find that $\sigma'^z_{xx} = \sigma'^z_{yy}$, and $\sigma'^z_{xy} = \sigma'^z_{yx} = 0$.

For bilayer ferroelectric WTe$_2$ with C$_{1v}$ symmetry, the BPVE tensor is

$$\left[\sigma^i_{jk}\right]_{3\times 6} = \begin{bmatrix} 0 & 0 & 0 & 0 & \sigma^x_{xz} & \sigma^x_{xy} \\ \sigma^y_{xx} & \sigma^y_{yy} & \sigma^y_{zz} & \sigma^y_{yz} & 0 & 0 \\ \sigma'^z_{xx} & \sigma'^z_{yy} & \sigma^z_{zz} & \sigma^z_{yz} & 0 & 0 \end{bmatrix}. \qquad \text{S(41)}$$

Due to lower symmetry, the BPVE tensor of bilayer WTe$_2$ is more complicated than those of bilayer MoS$_2$ [Eq. S(31)]. We find there are three independent in-plane BPVE coefficients $\sigma^x_{xy}$, $\sigma^y_{xx}$, $\sigma^y_{yy}$, and $\sigma^x_{xx} = \sigma^x_{yy} = \sigma^y_{xy} = 0$ due to $\widehat{M}_{yz}$ symmetry, and two independent out-of-plane BPVE coefficients $\sigma'^z_{xx}$ and $\sigma'^z_{yy}$ ($\sigma'^z_{xy} = \sigma'^z_{yx} = 0$) for ferroelectric bilayer WTe$_2$, which is consistent with our calculation results.

## Supplementary Note 4: In-plane and out-of-plane BPVE with the polarization of light

Now, we study the in-plane and out-of-plane BPVE response with the polarization of light. For a linearly polarized light, $\mathbf{E}(\omega)$ is a vector. For the 2D material under the normal incidence of light, $\mathbf{E}(\omega)$ only has the $E_x$ and $E_y$ components, and

$$\begin{cases} E_x = E\cos\theta, \\ E_y = E\sin\theta, \end{cases} \qquad \text{S(42)}$$

where $E$ is the magnitude of the light, and $\theta$ is the azimuthal angle relative to the $x$ axis.

According to Eq. S(31), the in-plane BPVE current of bilayer MoS$_2$ is

$$\begin{cases} j_x = -\sigma^y_{yy} E^2 \sin 2\theta, \\ j_y = -\sigma^y_{yy} E^2 \cos 2\theta. \end{cases} \qquad \text{S(43)}$$

If the light is the natural light that contains all-directions linearly polarized lights, $\theta$ varies from 0 to π, so the in-plane BPVE current is the sum of all the polarized light:

$$\int_0^\pi j_x = 0, \quad \int_0^\pi j_y = 0. \qquad \text{S(44)}$$

Therefore, the light should be linearly polarized to get the in-plane BPVE. The out-of-plane BPVE (light-induced polarization) of bilayer MoS$_2$ is

$$p_z = \left(E^2 \sigma'^z_{xx} \cos^2\theta + E^2 \sigma'^z_{yy} \sin^2\theta\right) = \sigma'^z_{xx} E^2, \qquad \text{S(45)}$$

because $\sigma'^z_{xx} = \sigma'^z_{yy}$, and $\sigma'^z_{xy} = \sigma'^z_{yx} = 0$. According to Eq. S(36), the out-of-plane BPVE polarization is isotropic and independent of the direction of the polarization of light, as shown in Fig. 2(f) in the main text.

Similarly, for bilayer WTe$_2$, we find in-plane BPVE current obeys

$$\begin{cases} j_x = \sigma^x_{xy} E^2 \sin 2\theta, \\ j_y = E^2 \left( \sigma^y_{xx} \cos^2 \theta + \sigma^y_{yy} \sin^2 \theta \right). \end{cases} \quad \text{S(46)}$$

The in-plane BPVE only shows the $\widehat{M}_{yz}$ symmetry, as shown in the main text. Because $\sigma'^z_{xy} = \sigma'^z_{yx} = 0$, $\sigma'^z_{xx} \neq \sigma'^z_{yy} \neq 0$, the out-of-plane BPVE polarization is

$$p_z = \sigma'^z_{xx} E^2 \cos^2 \theta + \sigma'^z_{yy} E^2 \sin^2 \theta. \quad \text{S(47)}$$

According to Eq. S(38), the light-induced BPVE polarization is anisotropic and dependent on the direction of the polarization of light, as shown in Fig. 3(f) in the main text.

## Supplementary Note 5: In-plane and out-of-plane BVPE of MoS$_2$ with interlayer sliding displacement

To show the dependence of in-plane and out-of-plane BPVE with interlayer sliding displacement, we manually change the interlayer sliding displacement defining as a parameter $\zeta$. $\zeta=0$ corresponds to no interlayer sliding, while $\zeta=+1/-1$ corresponds to the +P/–P phase.

In-plane and out-of-plane BVPE of bilayer MoS$_2$ with $\zeta$ are shown in **Fig. S10**. The in-plane BVPE coefficient is even with interlayer sliding displacement parameter $\zeta$, while the out-of-plane BVPE coefficient is odd with $\zeta$. The out-of-plane BVPE vanishes when $\zeta=0$, while the in-plane BVPE still exists. The in-plane BPVE is robust with the interlayer sliding displacement $\zeta$, while the out-of-plane BPVE is sensitive to the interlayer sliding displacement $\zeta$.

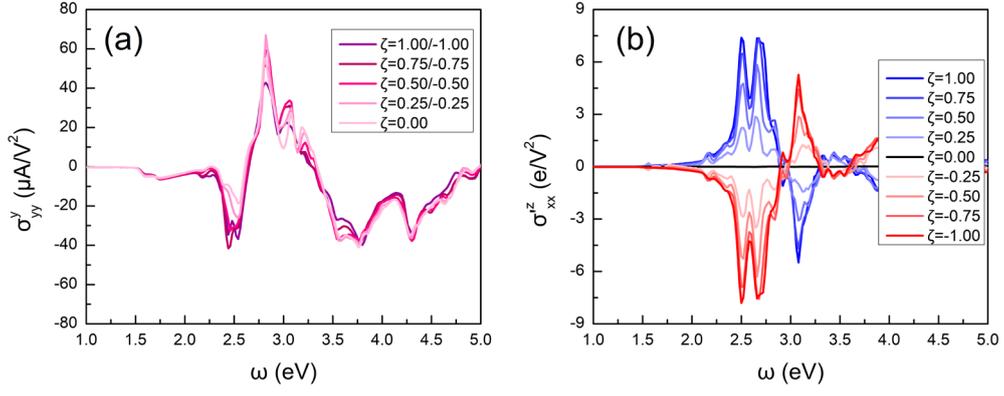

Fig. S10  (a) In-plane BVPE coefficient $\sigma_{yy}^{y}$ and (b) out-of-plane BVPE coefficient $\sigma_{xx}^{\prime z}$ of bilayer MoS$_2$ with interlayer-sliding displacement ζ.